# PoissonMat: Remodeling Matrix Factorization using Poisson Distribution and Solving the Cold Start Problem without Input Data

Hao Wang
Haow85@live.com
Ratidar.com
Beijing, China

*Abstract*—Matrix Factorization is one of the most successful recommender system techniques over the past decade. However, the classic probabilistic theory framework for matrix factorization is modeled using normal distributions. To find better probabilistic models, algorithms such as RankMat, ZeroMat and DotMat have been invented in recent years. In this paper, we model the user rating behavior in recommender system as a Poisson process, and design an algorithm that relies on no input data to solve the recommendation problem and the cold start issue at the same time. We prove the superiority of our algorithm in comparison with matrix factorization, random placement, Zipf placement, ZeroMat, DotMat, etc.

I. INTRODUCTION AND RELATED WORK

Recommender system is a lucrative business in modern day world. Companies such as Amazon, TikTok, etc. are spending lavishly on the technology to boost their traffic volume and revenues. Mainstream recommender system technologies focus on improvement of technical accuracy of the system, while in recent years, more people start to focus on other aspect of the research field such as fairness and cold-start problem.

The technical accuracy of recommender systems is important because for large websites, a slight increase in the accuracy metric leads to big revenue increase and dramatic marketing expenditure saving. Hiring a recommendation team with the necessary spending on computing facilities only costs a small share of the overall reduction in marketing costs. In fact, recommender system is one of a must-have products for large internet corporations.

In order to increase accuracy, researchers have invented algorithms such as collaborative filtering [1][2], matrix factorization [3], factorization machines [4], learning to rank [5][6] and deep learning models [7][8] since 1990's. Shallow models emerged first as the mainstream practice in the industry and they aimed to optimize technical metrics such as RMSE and MAE. Since the early 2010's, with the introduction of learning to rank, researchers' focus has shifted from accuracy metrics to ranking metrics, which instead of minimizing rating errors, emphasizes the preservation of the correct ranking order among recommendation items.

One of the major driving forces behind the technical evolution of recommender systems is the need to find better functional model to fit the user item rating data. The early day matrix factorization model is built as a probabilistic model using normal distributions. Then a more generic model named SVDFeature [9] was invented to make the exponent of the normal distribution as a weighted linear function of features. In 2021, Wang invented an algorithm named RankMat [10] that uses the power law family instead of the exponential family to fit the user item rating data. Later, ZeroMat [11] and DotMat [12] were proposed to furture utilize power law phenomenon to solve the recommender system in the cold-start context, i.e., when we do not have historic information of the new user / item.

In this paper, we use Poisson distribution to model the user-item rating behavior. We find the user-item-rating behavior modeled with Poisson distribution can better capture the interaction between the user and the item set. To make the model even better, we propose a version of the algorithm called PoissonMat that relies on no input data and solves the technical accuracy and cold-start problem at the same time.

The statistical framework behind matrix factorization is first introduced as the Probabilistic Matrix Factorization [13] . In this way, the probability of user rating given parameters is modeled by normal distribution. In the mean time, a framework named SVDFeature [9] was proposed as one of the most generic frameworks for matrix factorization that models the user / item feature vectors as linear combinations of features.

Other matrix factorization variants include timeSVD [14] , SVD++ [15] , LDA [16] , Alternating Least Squares [17] etc.

Each of the variants is designed for a specific scenario, but every one of them is proposed to increase the technical accuracy of the system.

In recent years, MatRec [18], Zipf Matrix Factorization [19] and KL-Mat [20] have been proposed to solve the fairness problem of recommender system, while ZeroMat [11] and DotMat [12] were introduced to solve the accuracy and cold-start problem at the same time. ZeroMat and DotMat are particularly important to the research in this paper. Both algorithms remodel the probabilistic framework of matrix factorization with power law distribution rather than normal distribution.

## II. PROBABILISTIC FRAMEWORK

The classic matrix factorization framework is to optimize the following formula :

$$L = \sum_{i=1}^{n}\sum_{j=1}^{m}\left(R_{i,j} - U_i^T \cdot V_j\right)^2 \quad (1)$$

, where $R_{i,j}$ is the user rating of the i-th user on the j-th item. U represents the user feature vector, and V represents the item feature vector.

To rebuild the classic matrix factorization model using probabilistic framework. We resort to the following framework:

$$P(U, V \mid R, \sigma_U, \sigma_V) \sim P(R \mid U, V, \sigma_U, \sigma_V) P(U, V \mid \sigma_U, \sigma_V) P(\sigma_U) P(\sigma_V) \quad (2)$$

In our PoissonMat framework, we consider R as a rare event that a user rates an item, and defines the probability of the event as follows :

$$P(R \mid U, V, \sigma_U, \sigma_V) = \frac{\lambda^{\frac{1}{rank}} \times e^{-\lambda}}{\left(\frac{1}{rank}\right)!} \quad (3)$$

, where $\lambda$ is a parameter, and is defined below :

$$\lambda = E(R) \sim \frac{1}{n}\sum_{j=1}^{n} R_{i,j} \quad (4)$$

By Zipf's Law, we have :

$$P(U, V \mid \sigma_U, \sigma_V) \sim U_i^T \cdot V_j \quad (5)$$

Plugging in Formulas (4) and (5) into Formula (3), we obtain :

$$P(U, V \mid R, \sigma_U, \sigma_V) \sim \frac{\left(\frac{1}{n}\sum_{j=1}^{n} U_i^T \cdot V_j\right)^{U_i^T \cdot V_j} e^{-\frac{1}{n}\sum_{j=1}^{n} U_i^T \cdot V_j}}{\left(\frac{1}{rank}\right)!}\left(U_i^T \cdot V_j\right) \quad (6)$$

We omit the prior probability of standard deviations for simplicity. After taking natural log of Formula (6), we obtain the following loss function :

$$L = \left((U_i^T \cdot V_j + 1)\log(U_i^T \cdot V_j) - U_i^T \cdot V_j\right) - \log\left(\left(\frac{1}{rank}\right)!\right) \quad (7)$$

Applying Stochastic Gradient Descent (SGD) algorithm to Formula (7), we acquire the following update rules for parameters U and V :

$$\frac{\partial L}{\partial U_i} = \left(\frac{(U_i^T \cdot V_j + 1)}{U_i^T \cdot V_j} + \log(U_i^T \cdot V_j) - 1\right) V_j \quad (8)$$

$$\frac{\partial L}{\partial V_j} = \left(\frac{(U_i^T \cdot V_j + 1)}{U_i^T \cdot V_j} + \log(U_i^T \cdot V_j) - 1\right) U_i \quad (9)$$

From Formulas (8) and (9), we find that the SGD update rules do not require any knowledge of the input data. However, to make the algorithm more practical and free of gradient explosion problem, we normalize the U and V vector in our computational procedures. In the end, we reconstruct the predicted unknown user item rating value by the following formula :

$$R_{i,j} = R_{max} \times U_i^T \cdot V_j \quad (10)$$

Just like ZeroMat and DotMat, PoissonMat is a new cold-start solution for recommender system that builds its theory upon a series of probabilistic approximation and simplification and needs no input data.

In addition to PoissonMat, we propose a hybrid model named PoissonMat Hybrid that takes PoissonMat as a data preprocessing step followed by classic matrix factorization. In the Experiment section, we demonstrate the superiority of both PoissonMat and PoissonMat Hybrid in comparison with 6 other algorithms.

The framework of PoissonMat is defined as :

---

Function PoissonMat :

1. Sample user id's from input user item rating matrix. Name the user id set U.

2. Sample item id's from the sampled user ids' rated item list. Name the item id set $I_i$, where i is the user id.

3. Initialize all the user feature vectors $\widehat{U}$ and feature vectors $\widehat{V}$ for the entire user id and item id sets to be random vectors.

4. for $U_i$ in U :
      for $V_j$ in $I_i$ :
        for iter_no in 1: max_iteration_number :
   $$U_i \mathrel{-}= \gamma \times \left(\frac{(U_i^T \cdot V_j + 1)}{U_i^T \cdot V_j} + \log(U_i^T \cdot V_j) - 1\right) V_j$$
   $$V_j \mathrel{-}= \gamma \times \left(\frac{(U_i^T \cdot V_j + 1)}{U_i^T \cdot V_j} + \log(U_i^T \cdot V_j) - 1\right) U_i$$

5. for i in user id list :
      for j in item id list :
   $$R_{i,j} = R_{max} \times U_i^T \cdot V_j$$

## III. DISCUSSION

The statistical framework behind the classic matrix factorization is the Probabilistic Matrix Factorization model. Since the method was proposed pretty early, both its derivation and underlying principles are pretty straightforward and simple. The philosophy behind PoissonMat is to use Poisson Distribution, instead of normal distribution to model the user rating behavior. The model is closer to the reality of human behavior than normal assumption.

The idea of applying more realistic functionals to better fit training data is the major theme of recommender systems, and artificial intelligence in general. The modern idea to do the task is using deep neural networks. As well known in the community, deep neural networks can approximate nearly any function in theory, deep learning engineering at its core is actually tweaking the functional space represented by neural networks to find the optimal functional that best fits the training dataset.

However, deep neural network design is something needs more scientific principles and explainability. In this paper, just like what we did in ZeroMat and DotMat, we rely on our mathematical knowledge, rather than trial and error in NAS-type research, to come up with a simplified version of matrix factorization based on Poisson distribution. Our approach, when compared with modern deep neural networks, may seem pre-stone-age dinosaurs. But we understand the underlying principles and mechanism behind our model much better than we do with deep neural networks.

Once again, just like ZeroMat and DotMat, we are facing the following question: Why a cold-start solution without input data is competitive with classic matrix factorization ? In another paper published by the author, he discussed the social impact of cold-start solution problem like ZeroMat - we are able to predict the preferences of individuals in our society with fairly accurate precision without knowing this individual's past. This implies there is an underlying mechanism behind human civilization that locks up people's preferences into a locked state. We need government interferences to break up this evolutionary trend in many social scenarios.

## IV. EXPERIMENT

We compare PoissonMat with the following algorithms: Random Placement, Zipf Placement, Classic Matrix Factorization, ZeroMat, ZeroMat Hybrid, DotMat, and DotMat Hybrid. We review these algorithms first before we demonstrate our experiment results.

### A. Random Placement

Random placement is the technique that recommends random items to users using uniform distribution. It is the most primitive algorithm that solves the cold-start problem since it requires no information of user / item data.

### B. Zipf Placement

It is well-known recommender system data is affected by popularity bias, so instead of recommending items to users using uniform data, we resort to Zipf distribution. We heuristically select the parameters for Zipf distribution, and recommend items to users using the distribution as the underlying probability distribution.

### C. Classic Matrix Factorization

The classic matrix factorization aims to minimize the following loss function :

$$L = \sum_{i=1}^{n} \sum_{j=1}^{m} \left(R_{i,j} - U_i^T \cdot V_j\right)^2 \quad (11)$$

The algorithmic problem can be solved using optimization approaches such as Stochastic Gradient Descent (SGD) , which uses random samples of the input data, rather than the whole data corpus, to compute the parameters. Matrix factorization can be solved fairly quickly due to the sampling method in SGD.

### D. ZeroMat

ZeroMat aims to solve the cold-start problem of recommender system, and it requires no input data. ZeroMat takes advantage of the Probabilistic Matrix Factorization framework and replaces normal distribution with dot product of user and item feature vectors based on the Zipf distribution theory.

The loss function of ZeroMat is defined as follows :

$$P(U, V \mid R, \sigma_U, \sigma_V) = \prod_{i=1}^{N} \prod_{j=1}^{M} (U_i^T \cdot V_j) \times \prod_{i=1}^{N} e^{-\frac{U_i^T \cdot U_i}{2\sigma_U^2}} \times \prod_{j=1}^{M} e^{-\frac{V_j^T \cdot V_j}{2\sigma_V^2}} \quad (12)$$

Like the classic matrix factorization, ZeroMat can be solved using SGD (The author simplified the formulas by setting standard deviations to be a constant number)   :

$$U_i = U_i + \gamma \times \left(\frac{V_j}{U_i^T \cdot V_j} - 2 \times U_i\right) \quad (13)$$

$$V_j = V_j + \gamma \times \left(\frac{U_i}{V_j^T \cdot U_i} - 2 \times V_j\right) \quad (14)$$

ZeroMat is proved to be superior than other cold-start heuristics, and sometimes is even competitive with the classic matrix factorization which has access to historic data.

ZeroMat does not demand any input data, but produces effective recommendation results.

## E. DotMat

DotMat borrows ideas from RankMat and ZeroMat. RankMat dismisses the exponential family as the backbone underlying probabilistic distribution for matrix factorization and proposes using power law distribution instead. Inspired by RankMat, DotMat also uses power law as the underlying probabilistic function, and defines the loss function as follows :

$$L = |(U_i^T \cdot V_j)^{U_i^T \cdot v_j} - \frac{R_{i,j}}{R_{max}}| \quad (15)$$

The optimal values of U and V are obtained using the following update rules :

$$U_i = U_i - \gamma((U_i^T \cdot V_j)^{U_i^T \cdot V_j} \text{sign}((U_i^T \cdot V_j)U_i^T \cdot V_j - U_i^T \cdot V_j)(1 + \log(U_i^T \cdot V_j))V_j) \quad (16)$$

$$V_j = V_j - \gamma((U_i^T \cdot V_j)^{U_i^T \cdot V_j} \text{sign}((U_i^T \cdot V_j)_i^T \cdot V_j - U_i^T \cdot V_j)(1 + \log(U_i^T \cdot V_j))U_i) \quad (17)$$

After acquisition of the optimal parameters, we reconstruct the unknown user item rating values using the following formula :

$$R_{i,j} = R_{max} \times (U_i^T \cdot V_j) \quad (18)$$

DotMat is proved to be superior to algorithms such as ZeroMat in real world environments.

## F. DotMat Hybrid

Analogous to ZeroMat Hybrid, DotMat Hybrid uses DotMat as a data preprocessing step to the ensuing matrix factorization algorithm. DotMat Hybrid outperforms all the afore-mentioned algorithms in open-sourced data sets.

To compare PoissonMat with the 6 algorithms we introduced in this section, we use Grid Search on the gradient learning steps (We notice very small steps achieve better results in our experiments, so we do grid search on the scale of $10^{-6}$). and test the algorithms on MovieLens 1 Million dataset [21] and LDOS-CoMoDa dataset [22].

Fig. 1 and Fig.2 illustrate the experimental results on MovieLens 1 Million Dataset. PoissonMat and PoissonMat Hybrid are competitive with DotMat and DotMat Hybrid. All these 4 algorithms are comparable with the classic matrix factorization algorithm which consumes millions of historic data records.

Fig. 3 and Fig. 4 demonstrate the experimental results on LDOS-CoMoDa dataset. With careful observation, we can safely draw the conclusion that PoissonMat and PoissonMat Hybrid produce some of the best results on a large sub-interval of parameter spectrum.

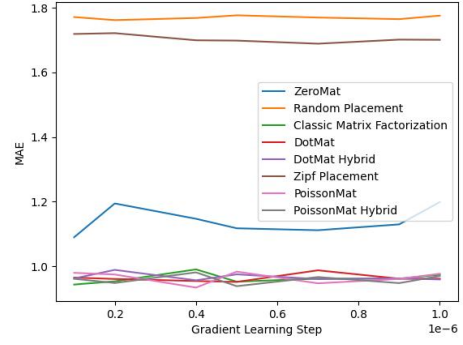

Fig. 1 Accuracy Comparison of different algorithms on MovieLens 1 Million dataset

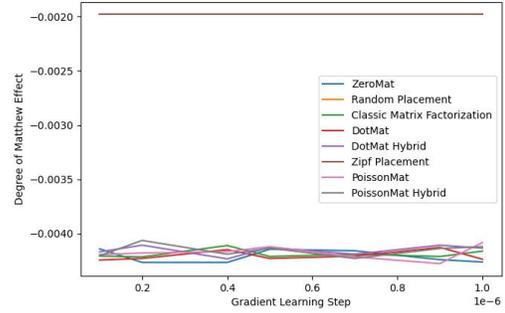

Fig. 2 Fairness Comparison of different algorithms on LDOS-CoModa dataset

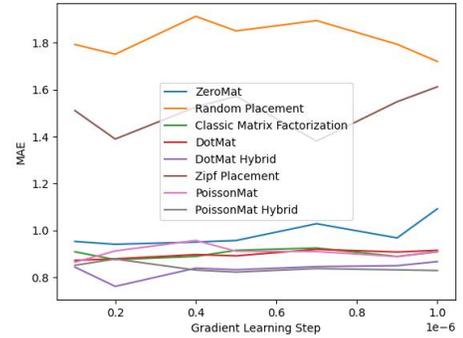

Fig. 3 Accuracy Comparison of different algorithms on LDOS-CoModa dataset

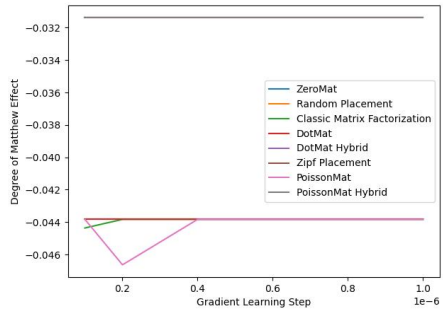

Fig. 4 Fairness Comparison of different algorithms on LDOS-CoModa dataset

## V. CONCLUSION

In this paper, we propose a new recommender system technology named PoissonMat and its hybrid version PoissonMat Hybrid. The new algorithm models the user rating behavior as a Poisson process instead of normal distribution as in the classic matrix factorization, or Zipf distribution as in ZeroMat.

In future work, we would like to explore the possibility of extending the cold-start problem solution into fields other than recommender systems.

REFERENCES

[1] B. Sarwar, G. Karypis, et. al. "Item-based collaborative filtering recommendation algorithms", WWW, 2001
[2] Z. Zhao, M. Shang, "User-Based Collaborative-Filtering Recommendation Algorithms on Hadoop", International Conference on Knowledge Discovery and Data Mining, 2010
[3] Y. Wang, Y. Zhang, "Nonnegative Matrix Factorization: A Comprehensive Review", IEEE Transactions on Knowledge and Data Engineering, 2013
[4] S. Rendle, "Factorization Machines", ICDM, 2010
[5] M. Morik, A. Singh, J. Hong, T. Joachims. "Controlling Fairness and Bias in Dynamic Learning-to-Rank", SIGIR, 2020
[6] H. Yadav, Z. Du, T. Joachims. "Fair Learning-to-Rank from Implicit Feedback", SIGIR, 2020
[7] H. Guo, R. Tang, et. al., "DeepFM: A Factorization-Machine based Neural Network for CTR Prediction", IJCAI, 2017
[8] H. Xue, X. Dai, J. Zhang, et. al. "Deep Matrix Factorization Models for Recommender Systems", Proceedings of the Twenty-Sixth International Joint Conference on Artificial Intelligence, 2017
[9] T. Chen, W. Zhang, et. al. "SVDFeature: A Toolkit for Feature-based Collaborative Filtering", Journal of Machine Learning Research, 2012
[10] H. Wang, "RankMat: Matrix Factorization with Calibrated Distributed Embedding and Fairness Enhancement", ICCIP, 202
[11] H. Wang, "ZeroMat: Solving Cold-start Problem of Recommender System with No Input Data", IEEE 4th International Conference on Information Systems and Computer Aided Education, 2021
[12] H. Wang, "DotMat: Solving Cold-start Problem and Alleviating Sparsity Problem for Recommender Systems", ICET, 2022
[13] R. Salakhutdinov, A. Mnih, "Probabilistic Matrix Factorization", NIPS, 2007
[14] Y. Koren, "Collaborative Filtering with Temporal Dynamics", KDD, 2009
[15] Y. Koren, "Factorization meets the neighborhood: a multifaceted collaborative filtering model", KDD, 2008
[16] Z. Tang, X. Zhang, et. al. "LDA Model and Network Embedding-Based Collaborative Filtering Recommendation", International Conference on Dependable Systems and Their Applications (DSA), 2019
[17] G. Takacs, D. Tikk. "Alternating Least Squares for Personalized Ranking", The 6th ACM Conference on Recommender Systems", 2012
[18] H. Wang, "MatRec: Matrix Factorization for Highly Skewed Dataset", ICBDT, 2020
[19] H. Wang, "Zipf Matrix Factorization: Matrix Factorization with Matthew Effect Reduction", ICAIBD, 2021
[20] H. Wang, "KL-Mat: Fair Recommender System via Information Geometry", icWCSN, 2022
[21] T. Bertin-Mahieux, B. Whitman, P. Lamere, The Million Song Dataset, ISMIR, 2011
[22] ODIĆ, Ante, TKALČIČ, Marko, TASIČ, Jurij F., KOŠIR, Andrej: Predicting and Detecting the Relevant Contextual Information in a Movie-Recommender System, Interacting with Computers, Volume 25, Issue 1, 2013